# The glocal forest

Efrat Seri[1*], Elad Shtilerman[2*] and Nadav M. Shnerb[1]

1. Department of Physics, Bar-Ilan University Ramat-Gan 52900, Israel.

2. Porter School of Environmental Studies, Tel Aviv University, Israel.

∗ These authors contributed equally to this work.

Ecological spatial patterns reflect the underlying processes that shape the structure of species and communities. Mechanisms like inter and intra species competition, dispersal and host-pathogen interactions are believed to act over a wide range of scales, and the inference of the process from the pattern is, despite its popularity, a challenging task.  Here we call attention to a quite unexpected phenomenon in the extensively studied tropical forest at the Barro-Colorado Island (BCI): the spatial deployment of (almost) all tree species is statistically equivalent, once distances are normalized by $\ell_0$, the typical distance between neighboring conspecific trees. Correlation function, cluster statistics and nearest-neighbor distance distribution become species-independent after this rescaling.  Global observables (species frequencies) and local spatial structure appear to be interrelated. This "glocality" suggests a radical interpretation of recent experiments that show a correlation between species' abundance and the negative feedback among conspecifics. For the forest to be glocal, the negative feedback must govern spatial patterns over *all* scales.

# Introduction

An understanding of the forces that govern the dynamics of populations and communities is one of the major challenges of contemporary ecology [1]. Besides its practical importance for management and conservation, the apparent excess biodiversity in systems like tropical forests, coral reef and freshwater plankton poses a thought-provoking conceptual problem, as it seems to violate the competitive exclusion principle [2]. The numerous explanations suggested to this puzzle [3-9], and the hot debates around them, just reveal that a clear understanding of these mechanisms is still lacking.

One of the most useful probes for these systems, and in particular for communities of sessile species, is the statistical properties of their spatial patterns. This feature may reflect dispersal limitations [10,11], competition [12-15], interactions with pathogens and predators [16], environmental filtering [17] and many other circumstances [18]. Accordingly, numerous analyses of aggregation, patchiness and structure have appeared during the last decade, many of them were utilizing the data from the spatially explicit, long-term monitoring of a few tropical forests provided by the Center for Tropical Forest Science (CTFS). Virtually all these works implemented various techniques of point pattern analysis [19], taking every tree as a point in space and retrieving quantities like the correlation function from the matrix of distances between individuals.

Here we report the results of a few point pattern analyses for >1cm trees and understory plant in the Barro-Colorado plot [20-22]. The only difference between our work and previous studies is that, for any given species, we have normalized the distances between trees by the species-specific fundamental length-scale $\ell_0$, defined via $\ell_0 = \sqrt{A/N_i}$, where A is the plot area and $N_i$ is the number of conspecific trees (or understory shrubs) in that area (i.e., the abundance of the i-th species). Amazingly, after this rescaling it appears that all species have a statistically identical spatial structure (at least up to the length scale of the plot), with minor differences that appear mainly below $\ell_0$. This outcome imposes a very strong constraint on the models that may explain or describe spatial distributions. Our results suggests that a reasonable model has to be glocal, i.e., to allow the overall abundance of a species (a global property) to dictate the local structure or vice versa.

A possible mechanistic interpretation of our results may be related to recent experiments of Mangan et. al. [23]. These authors have analyzed the negative feedback between conspecific trees for 6 species in the BCI forest, pointing out that the strength of this feedback varies among species and is a good predictor of the relative abundance of a species. Our results suggest a far-reaching generalization of these findings, namely, that the negative feedback is (for *almost all* species) the dominant factor in governing the statistical properties of spatial patterns. Other effects, and in particular interspecific interactions [15] and dispersal limitations [24], appear to be relatively weak.

## Results

Let us start with one of the standard measures of spatial structure, the nearest-neighbor distance distribution (NNDD) [25]. In Fig. 1 (left panels) we plot this quantity for the 15 most abundant species and for all species with more than 50 individuals in the forest. The statistics is quite good, and one observes, beyond the scale of a few meters, an exponential decay of the distribution over four orders of magnitude. $P(r)$, the chance to find the nearest neighbor tree at a distance $r$, follows (at $r > 0.5\ell_0$) $P(r) \sim \exp(-\alpha r)$ where, as expected, $\alpha$ is an increasing function of the abundance. Such an exponential decay in two spatial dimensions is very interesting by itself and its origin is not clear; in any case, the results clearly exclude a Gaussian decay of the NNDD (as predicted for a Poisson forest) and a crossover from Gaussian to power-law one finds for negative-binomial spatial distributions.

Here we would like to highlight another property: once the same graphs are plotted for the rescaled distance $r/\ell_0$ (Figure 1, right panels), the exponent $\alpha^* \equiv \alpha \ell_0$ becomes (almost) abundance independent for $N_i > 150$ species. The clear correlation between the exponent and the abundance practically disappears after rescaling, and the data collapse is quite impressive.

For $N_i < 150$, Fig. 1(f) indicates an average increase of $\alpha^*$ with abundance. This may be an artifact of the non-universal behavior at short length scales, and in any case it is difficult to assess the statistical significance of the results when the number of points is so small. However, if this effect is significant it suggests that, with respect to this measure, rare species are less aggregated than the frequent species, contrary to the conclusion of the very influential paper by Condit et. al. [26] and in agreement with [27].

To emphasize the novelty of this collapse, we show, in the supplementary material (section 2), the corresponding figures for four mechanistic models that were used in the literature to account for the spatial deployment of forests: A Poisson process, spatial neutral dynamics with mixed local-global recruitment kernel (MLGK), which is similar to the Cox process, spatial neutral dynamics with a Cauchy (fat-tailed) kernel and the fractal structure (random Cantor set) suggested in [28].

As shown in the supplementary, the Poisson forest indeed shows a data collapse in the rescaled coordinates (since it admits only one length scale which is determined by the abundance), but the function $P(r)$ is a Gaussian, not exponential, as expected for random point patterns [25]. The two processes with a finite kernel (MLGK and Cauchy) admit two scales: one is determined by the kernel, the other by the abundance. As a result, they both fail to yield a collapse and the plots of $\alpha$ and $\alpha^*$ vs. abundance show a clear trend. A fractal structure admits, again, a single length scale, but now this scale has nothing to do with the abundance (the basic scale may reflect, say, spatial heterogeneity), so the collapse occurs in the "wrong" plot, when the length scale is not normalized.

A similar effect of rescaling is revealed when the correlation function $g(r)$ is considered. This function (also known as $\Omega$-ring statistic, or the radial derivative of Ripley's K-function) was implemented in [26] to show that rare species in the tropical forest are more aggregated than common species. The results of [26] are illustrated again in figure 2(e), showing a general trend towards lower correlations for more frequent species. After rescaling (Fig. 2, right panels) this relation becomes non-significant. Since the decay of correlations above $r > 0.5\ell_0$ resemble a power-law (in agreement with [10,11,28]), the rescaling of length does not affect the slope. Still, the absolute height of the correlation function is independent of the abundance when the comparison is made at the same value of $r/\ell_0$.

The corresponding graphs in the supplementary (section 3) show that the mechanistic model cannot yield these features. In a Poisson forest the correlations are independent of the distance (up to noise and finite size effects), in the two dispersal models the low-abundance species appear to be more clustered, and the same property characterizes also the fractal forest.

Both NNDD and the correlation function have a limited ability to separate length scales: these are probability distribution functions, and the normalization condition dictates anti-correlations between their values at short and long distances. As pointed out recently by [29], one would like to decompose variance scale by scale. To address this requirement we provide a third piece of evidence in Fig. 3. Here the aggregation is characterized by the total number of tree clusters, with a varying grid scale $\ell$. The two-dimensional map of the plot is covered by a mesh of $\ell \times \ell$ squares, each square colored black if it contains at least one tree of the focal species. Two black squares are in the same cluster if they are connected by a path of nearest neighbor black sites (see a detailed description in [30,31]). The total number of different clusters, $F(\ell)$, approaches $N_i$ when $\ell \to 0$ and converges to 1 for large $\ell$. Counting the number of clusters at different spatial resolutions Figure 3 demonstrates, again, a quite good data collapse and vanishingly small correlation between the number of clusters and the species abundance in the rescaled data.

The figures in section 4 of the supplementary demonstrate again that none of the mechanistic models imitates the real forest. The Poisson forest shows, as expected, a data collapse, but it deviates strongly from the empirical results, as there is no real clustering in a Poisson forest, the decay of $F(\ell)$ is much slower. The fractal forest data collapses, if any, in the non-normalized diagram, and the two dispersal models fail to yield a collapse.

## Discussion

All the indicators we have analyzed here are pointing to a quite surprising feature: the spatial structure of all tree species (up to a few exceptions) is almost identical, when measured with respect to the species specific length scale $\ell_0$. Such a feature emerges trivially in a Poisson forest (if $N_i$ trees from the i-th species are randomly distributed in the forest, there is only one length scale $\ell_0 = \sqrt{A/N_i}$ and the spatial aggregation characteristics will become $N_i$-independent after rescaling by $\ell_0$) or in a "lattice" forest, where trees of every species are located on the vertices of a two-dimensional squared lattice with a lattice constant $\ell_0$.

However, these two models are irrelevant since the aggregation in the forest is well-known to be stronger than Poisson [26], let alone of a lattice. Indeed, all our parameters deviate substantially from the Poisson/lattice limits; still the forest admits only a single scale.

On the other hand, it seems that every non-Poissonian mechanistic model of forest dynamics must admit at least one typical, species specific length scale associated either with the recruitment kernel, i.e., with the distribution of distances between mother and offspring (encapsulating the dispersal kernel and the chance of a seedling to capture an open slot) or with spatial heterogeneity. Our results show that this "local" scale, whatever it is, dictates (or is dictated by) the "global" scale $\ell_0$ associated with the overall density of the focal species in the forest. This puts a severe restriction on the space of possible models and, in fact, the property of glocality is not a part of *any* of the models we are familiar with, from neutral dynamics [10,11] to tradeoffs [9] to niche models. In all these mechanistic theories the local dynamics has nothing, or almost nothing, to do with the overall abundance of a species.

Former results, like the excess positive correlations of rare species observed in [26], may indeed reflect the dominance of a single length scale. The correlation function $g(r)$ quantifies the information one has about the density fluctuations at a distance $r$, given the presence of a tree at the origin. If the rescaled measure $g(r/\ell_0)$ is roughly the same for all species, it implies that the information for frequent species (small $\ell_0$) falls faster on real scale, rendering the infrequent species more aggregated.

It was already pointed out in [26], the trees of some exceptional species form circular clumps, apparently reflecting the effect of dispersal limitation. The spatial patterns of these species, like *Rinorea sylvatica, Anaxagorea panamensis* and *Bactris major* (see spatial patterns in Supplementary 5) must admit at least two length scales – one associated with the cluster size, the other with the inter-patch distance. These exceptional species are not glocal, and indeed appear as outliers in Figs. 1-3. For example, the two irregular points (diamond markers) in Fig 1e correspond to *Rinorea sylvatica* and *Bactris major,* where the value of $\alpha$ for *Anaxagorea panamensis* is so high that it cannot be seen in the frame presented here. The same phenomenon is evident in Figures 2 and 3. Moreover, the correlation function for these species clearly reveals a crossover between two slopes, as opposed to the single power-law for "standard" species. Still, the number of exceptional species is relatively small.

We believe that this glocality is not a peculiarity of the BCI forest, but a characteristic of other forests as well. Unfortunately we were not successful in obtaining access to the spatially resolved data of the other homogenous forest in the CTFS system in Pasoh, Malaysia. Hopefully other groups will have the opportunity to compare the rescaled spatial structure in different locations. Anyhow, at least any model for the BCI data should allow a glocal interrelation between scales.

Apparently, glocality requires a species specific mechanism that carries information along scales. Such a mechanism may act either top-down (e.g., when the overall density controls the dispersal kernel since the typical movement of a species specific animal disperser is proportional to the distance to the nearest neighbor conspecific individual) or bottom-up, for example, by negative feedback which reduces the probability of conspecific seedlings to survive in the proximity of an adult tree [29].

Indeed, Mangan et. al. [23] have demonstrated experimentally this mechanism, attributing it to plant-soil feedback mediated by soil biota and further showed that the feedback is a good predictor of the abundance of different species in the plot. As pointed out in the introduction, glocality may emerge from this mechanism, provided that the dominant factor that controls spatial patterns is the strength of "repulsion" between conspecific trees. However, this is a strong requirement, as it implies that all other length scales, like those involved in dispersal limitation and inter-specific competition have only a minor effect on the spatial deployment of the forest.

If true, our findings and the work of [23] suggest a model in which every species has its own mutual exclusion radius. Such a model may also explain the global correlations in the abundance of taxonomic groups [32]. Still, the role of competition in shaping community assembly and spatial structure is not clear. One may imagine a model in which competitive superiority is balanced by stronger negative feedback [similar to the tradeoff models of [9] ] or neutral dynamics, but at this point we do not know under what conditions these dynamics do not alter the statistical properties of a spatial patterns.

**Acknowledgements**: We thank David Kessler, Lewi Stone, Joe Wright and Richard Condit for helpful discussions and comments. This work was supported by the Israeli Ministry of Science and Technology TASHTIOT program and by the Israeli Science Foundation grant no. 454/11 and BIKURA grant no. 1026/11. This research was supported by the Porter school of Environmental Studies at Tel Aviv University. The BCI forest dynamics research project was made possible by National Science Foundation grants to Stephen P. Hubbell: DEB-0640386, DEB-0425651, DEB-0346488, DEB-0129874, DEB-00753102, DEB-9909347, DEB-9615226, DEB-9615226, DEB-9405933, DEB-9221033, DEB-9100058, DEB-8906869, DEB-8605042, DEB-8206992, DEB-7922197, support from the Center for Tropical Forest Science, the Smithsonian Tropical Research Institute, the John D. and Catherine T. MacArthur Foundation, the Mellon Foundation, the Small World Institute Fund, and numerous private individuals, and through the hard work of over 100 people from 10 countries over the past two decades. The plot project is part the Center for Tropical Forest Science, a global network of large-scale demographic tree plots.

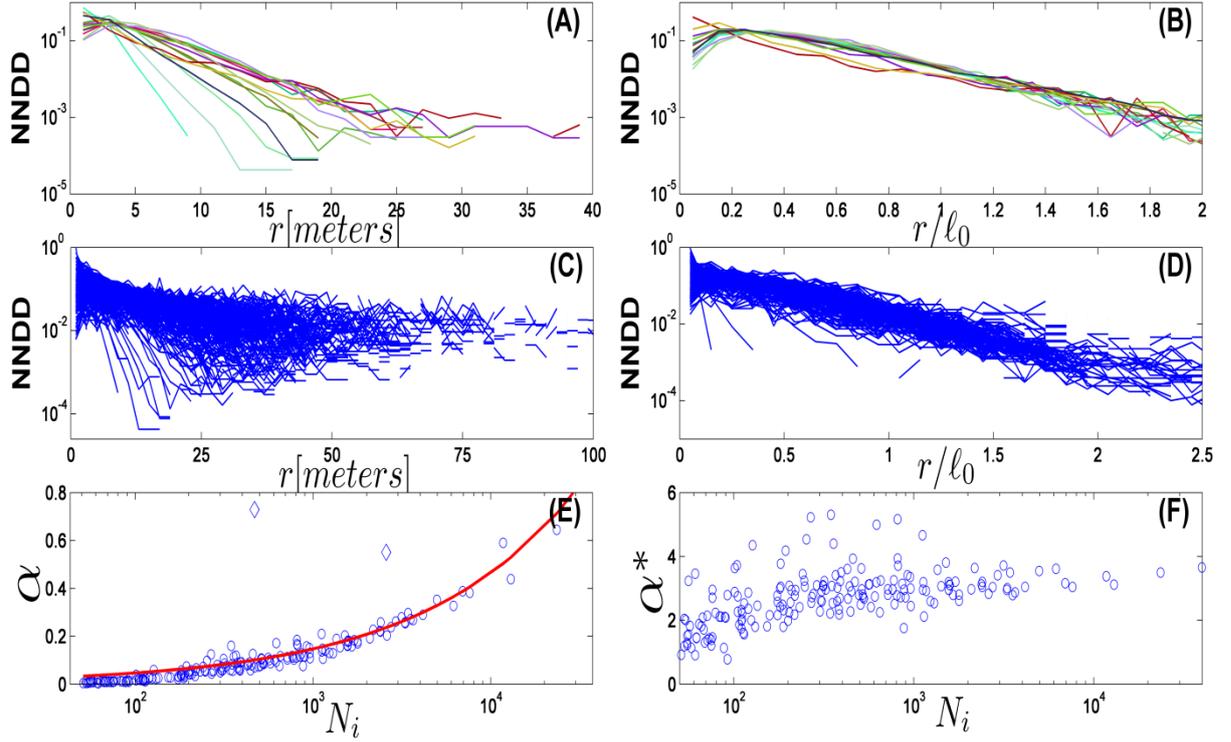

**Figure 1: Nearest neighbor distance distribution (NNDD) in the BCI tropical forest.** For every species, the graph presented is a normalized histogram of the distances from every tree to its closest conspecific tree in the forest [25]. In panel (A) the data is presented for the 15 most abundant species on a semi-log scale. The decay at long distances is clearly exponential, but the slope varies strongly among species. Panel (B) shows the same data when distances are rescaled by $\ell_0$, showing a good data collapse in the exponential regime. Panels (C) and (D) are the same graphs for all species with >50 individuals in the forest. The data is noisier, but the pronounced features are the same. In panel (E), the slope $\alpha$ in the exponential regime is plotted against the abundance of the species $N_i$, showing a $\sqrt{N_i}$ dependence as expected (red line, notice the logarithmic scale). This correlation is much weaker, and almost disappears for $N_i < 150$ when the rescaled data is analyzed (panel (F), $\alpha^*$ is the slope measured in rescaled coordinates).

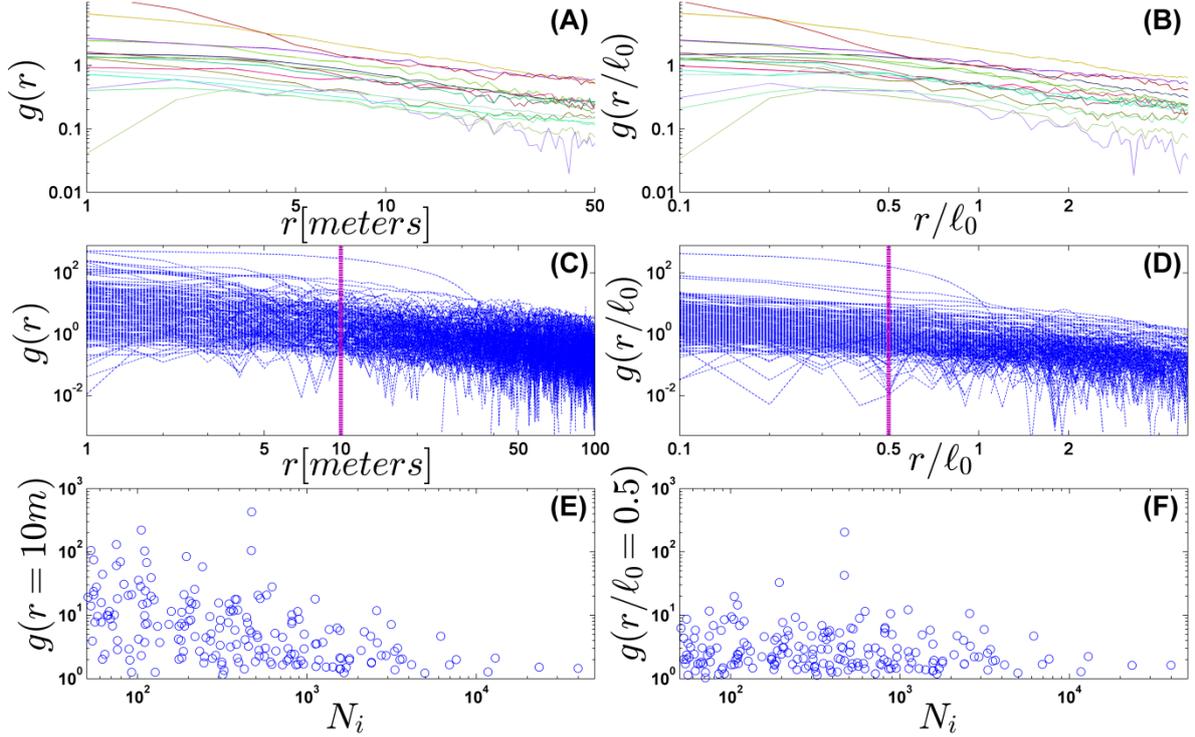

**Figure 2: The correlation function.** For each individual $i$, two quantities are measured: $s_i(r)$, the number of conspecific trees in a ring of inner distance (from the focal tree) $r$ and outer distance $r+\Delta r$, and $A_i(r)$, the fraction of this ring area that is included in the 50-ha plot. The correlation function $g(r) \equiv \dfrac{\sum s_i(r)}{N_i \sum A_i(r)} - 1$ (the sum runs over all focal-specie's trees) measures the information about density fluctuations embodied in the observation of a tree at $r=0$. $g(r) = \Omega(r) - 1$, where $\Omega$ is the relative neighborhood density defined in [26], and is zero for a homogenous Poisson process. At long distances the decay of the correlations is described quite faithfully by a power law [28] (panel A, 15 most abundant species, panel C, all species). The short-distance correlation tends to be higher when a species is rare. This property is demonstrated in panel (E), where the height of the correlation function at $r=10m$ is plotted against the abundance, in parallel with Fig. 2 of [26]. The rescaled plots (panels B and D) have the same slopes, since a power law is scale independent, but the crossover to the power law behavior is close to $r/\ell_0 = 1$ for almost all species. Moreover, the association between abundance and correlation becomes non-significant in the rescaled plots, as seen in panel (F), showing the height of $g(r/\ell_0 = 0.5)$ vs. $N_i$. Pearson correlation coefficient in panel (E) is 0.46, with p-value less than 0.01, while for (F) the coefficient is 0.12 and the p-value is 0.1.

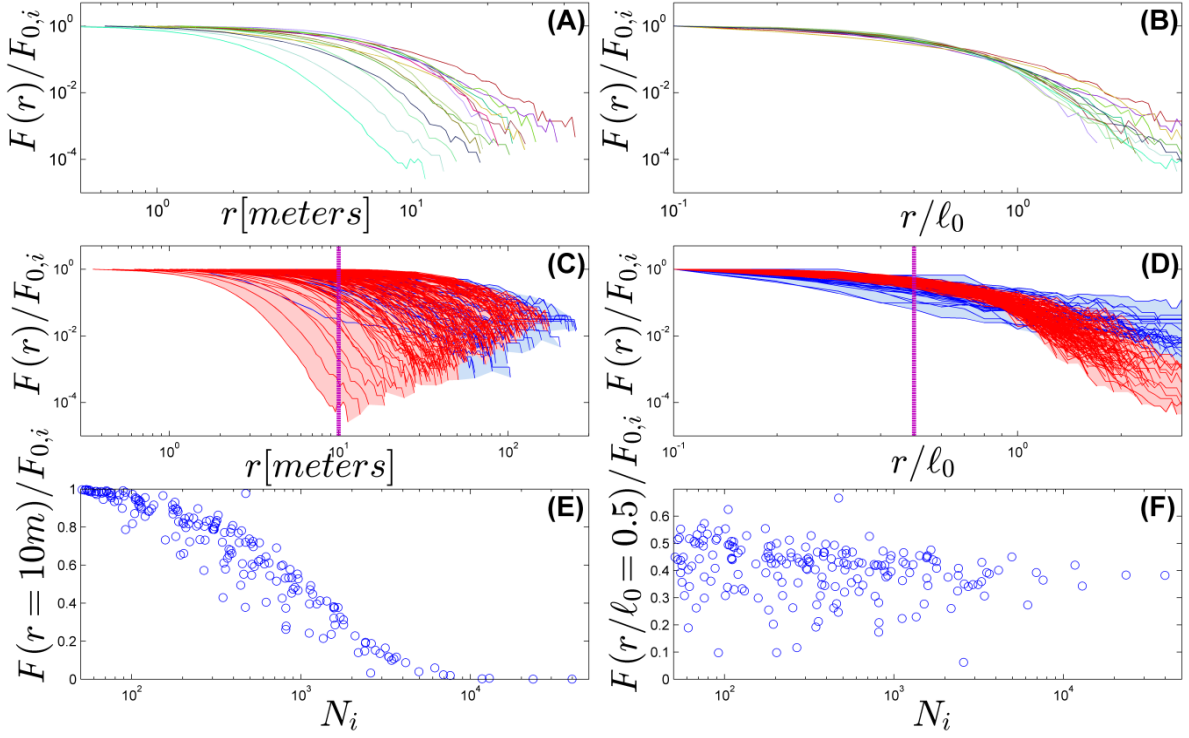

**Figure 3: Separating spatial scales using patch statistics**. The two dimensional 50-ha plot map was covered by a grid of $\ell \times \ell$ squares, every square is "black" if it contains at least one tree of the focal species, otherwise it is "white". $F(\ell)$, the number of clusters at each resolution level, is normalized by $F_{0,i}$, the number of clusters for the minimal value of $\ell$ used, $F_{0,i} \to N_i$ as $\ell \to 0$. Panels (A) and (C) show $F(\ell)/F_{0,i}$ vs. $\ell$ for the 15 most abundant species (A) and for all species with more than 50 individuals in the forest (C). Panel (E) depicts $F(\ell=10m)/F_{0,i}$ vs. $N_i$ (a cut along the purple dashed line in panel (C)), emphasizing the abundance dependence of the result. Panels (B) and (D) show the same analysis when $F(\ell)/F_{0,i}$ is plotted vs. $\ell/\ell_o$. The data collapses (we colored red 158 out of 193 species (82%), to emphasize that most of the width is due to a small number of exceptional species, see supplementary material 5) and the correlation with abundance disappears (panel F, showing $F(\ell=0.5\ell_o)/F_{0,i}$ vs. $N_i$).

# Supplementary material

## 1. The mechanical models and simulation procedures

The results presented in the main text were extracted from the spatial deployment of trees and undercanopy in the BCI forest (the results used in this paper are of the first census). Here we present results obtained using the same set of analyses, when applied to a few *simulated* forests, each represents a popular mechanistic model which is implemented in the literature in order to explain the spatial structure of forests and other systems. To check for a data collapse and its (in)dependence on abundance, we have simulated a single species dynamics for each of the models below, on a plot of size 500x1000m$^2$ (which is the size of the BCI plot), until the process reaches the prescribed abundance. The results for species with 40000, 20000, 10000, 4000, 2000, 1000 and 500 individuals were analyzed and compared. For some technical complications we have used a different set of numbers for the fractal forest (see 4 below), but there is no reason to think that it may change any general features of the results.

The models considered in this supplementary are:

1. **Poisson forest**. This is the simplest model, assuming that there is no spatial correlation between the mother tree and its offspring. Although the recruitment kernel must depend on the distance, this model becomes accurate when the linear size of the surveyed plot is much smaller than the typical length associated with the recruitment.

2. **MLGK (Cox-like) forest**: The Cox process is a result of a two-stage random mechanism. To build a Cox forest of N trees one choses m points (centers) at random, and place N/m trees (again at random) within a distance r from every center. To make the process slightly more realistic, we have implemented here a neutral dynamics with mixed local-global kernel, a model that we have used in a recent paper analyzing the spatial structure of the BCI plot. [11]. Starting with a single individual from the focal species, the neutral dynamics is implemented. In every elementary timestep two individuals are picked at random and the offspring of the (randomly selected) first replaces the second, see [33]. Once the first individual is picked, the second is chosen at random from its 2-meters neighborhood with probability 1-μ and from the whole plot with probability μ.

   The process continues until the desired number of individuals (the species abundance) is obtained, all other details are given in [11]. The limit μ=1 corresponds to the Poisson forest, for smaller μ-s every population is made of a random (Poisson) collection of clusters of individuals. In the simulation here we have used μ=0.1, as this value yielded the best (although unsatisfactory) fit to the BCI data in [11].

3. **Cauchy forest**: is generated using the same neutral dynamics algorithm [see again [11]], but now the recruitment kernel is Cauchy, i.e., the probability that the descendent of a tree feels a gap at a distance r is given by:

$$K(r) = \frac{1}{\pi\gamma}\frac{1}{1+r^2/\gamma^2}.$$

Here $\gamma$ is the characteristic spatial scale of the kernel. In this type of kernel, there is no specific distance that separates global dispersal from local; rather, the probability decreases slowly with r. In [11] we showed that the clusters obtained from a neutral process with this kernel fit quite nicely the BCI forest data.

4. **A Fractal** forest: to simulate a forest with a fractal structure, we have implemented the random Cantor set algorithm suggested as a model for tropical forests by (Green (2000)). Starting with a 2x2 array, each cell is chosen to be empty with probability P or is chosen to be "active" with probability 1-P. Each of the active cells is then divided into 4 equal squares and the process is iterated. The active sites of the last iteration are the focal species trees.

We have stopped the process when the forest reaches the size of 1024x1024 cells, using P=0.75. Implementing a few realizations of the same algorithm, we were able (due to the randomness of the process) to generate a few sets of focal species trees, sets that have the same fractal structure but different abundance. For the analysis presented below we have used realization with $n$ "trees" where $n \approx 40000$, 20000, 10000, 4000 and 2000.

In the following sections we show the results obtained when we applied our measures to the various simulated forests. The panels of every figure correspond to the two upper panels [(A) and (B)] and the two lower panels [(E) and (F)] of the figures shown in the main text.

## 2. Nearest-neighbor distance distribution (NNDD)

The NNDD for species with different abundance will give a data collapse when the distances are normalized by $\ell_0$ if this is the only length scale in the forest. This is clearly true for a Poisson process but Figure 2.1 indicates that P(r) is a *Gaussian* (this is a known feature of random point-patterns in two dimensions), not the exponential distribution that characterizes the empirical data.

For the MLGK model (Figure 2.2) one can observe the crossover to Poisson statistics at long distances, but the short distance data does not collapse, and the Cauchy simulation (Figure 2.3) do not have a collapse region at all. Finally, the NNDD for a fractal forest (Figure 2.4) does collapse, but this occurs in the non-normalized (left) graph, since all fractals have the same basic length scale.

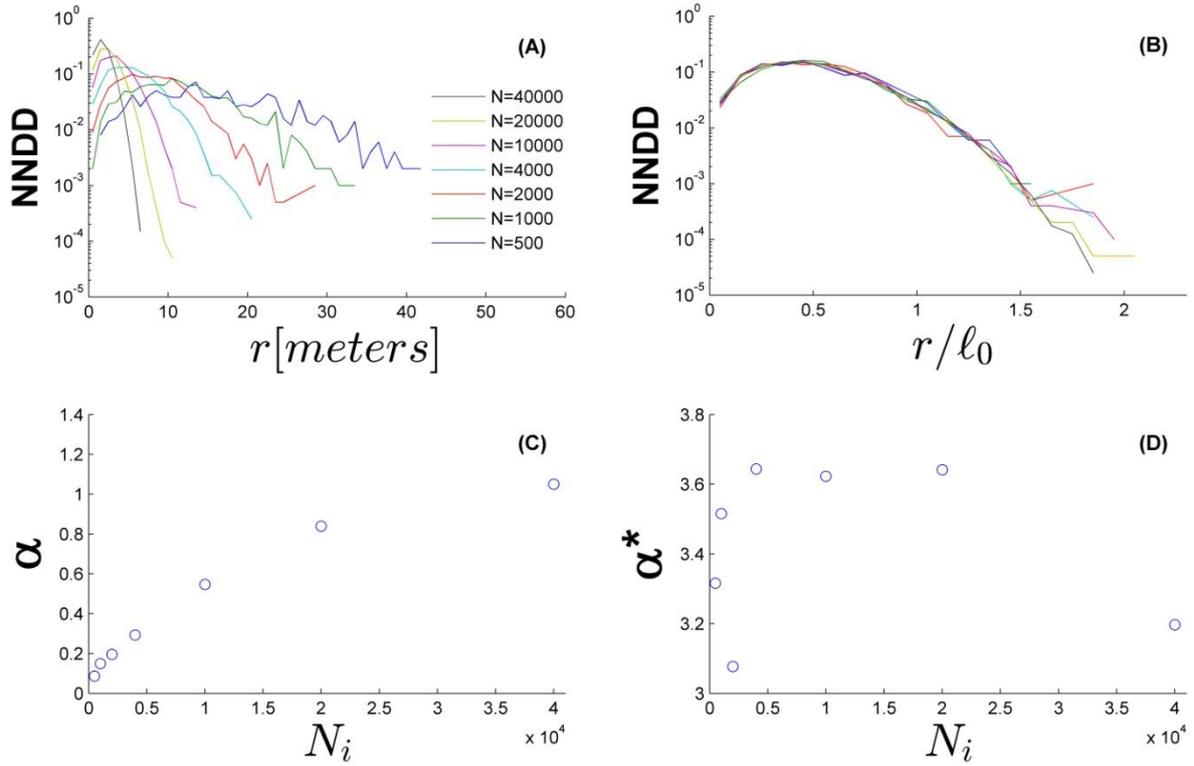

Figure 2.1: NNDD for Poisson distribution. (A) vs. the real distance r (to be compared with Fig. 1A of the main text). (B) vs. the normalized distances $r/\ell_0$ (to be compared with Fig. 1B of the main text). (C) The slope of NNDD as a function of the species abundance N for (A) (to be compared with 1E), (D) :same as (C) but for the normalized curves in (B) (in parallel with panel 1F in the main text). In panels (C) and (D) the "slope" of the tail was measured as if the decay is exponential, just to show the results that correspond to the analysis presented in the main text. Indeed, of course, the decay is Gaussian so there is no reasonable way to compare the two systems.

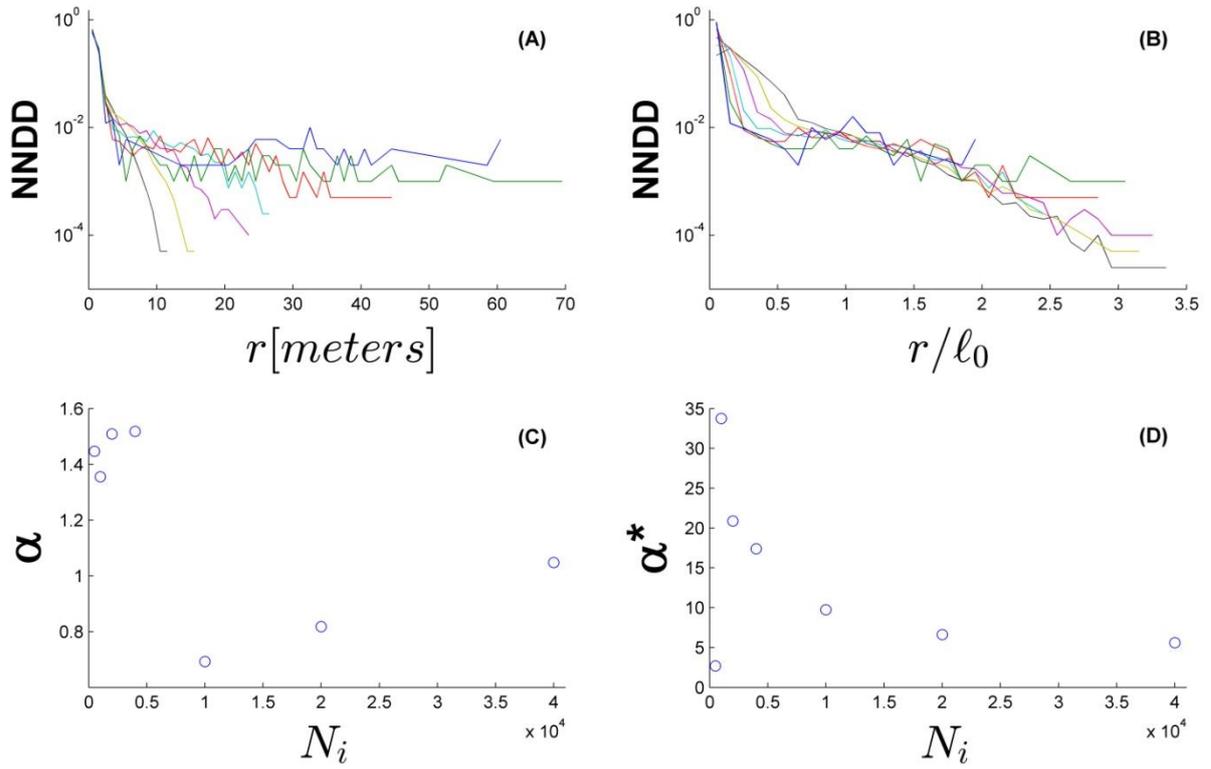

Figure 2.2: Same as fig. 2.1 but for MLGK (μ=0.1).

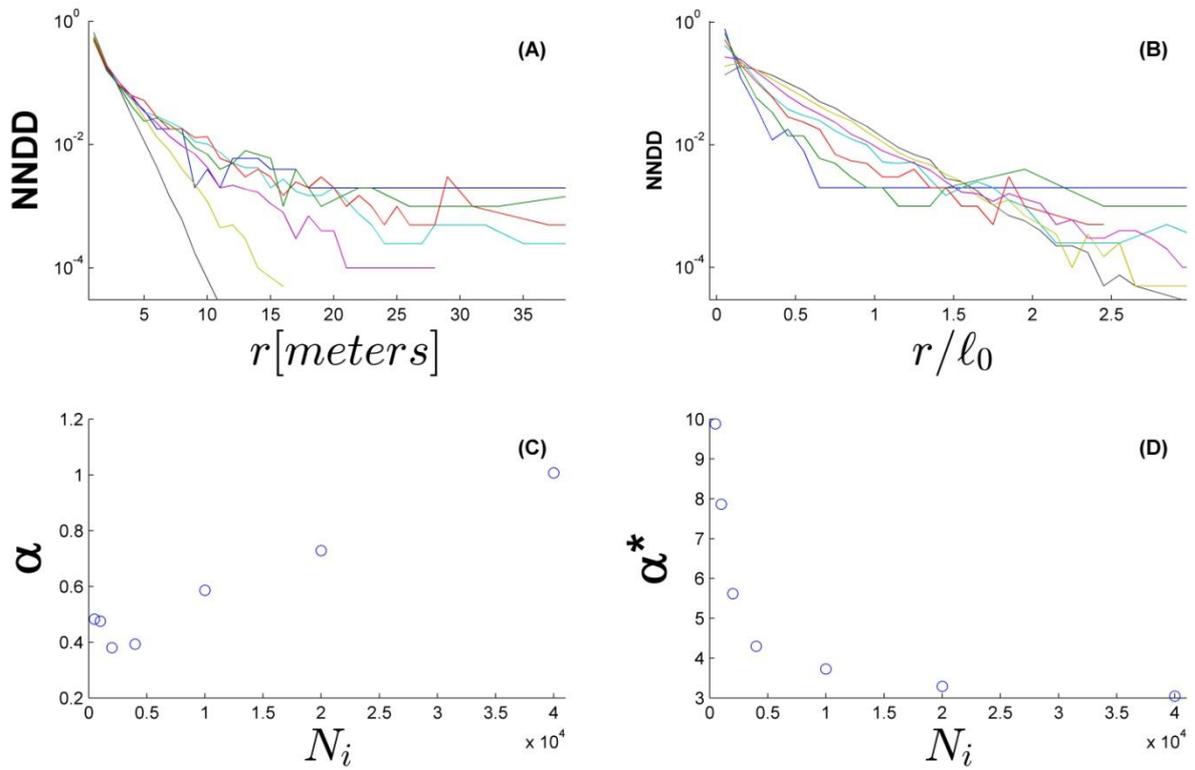

Figure 2.3: Same as fig. 2.1 here for the Cauchy kernel (ɣ=20)

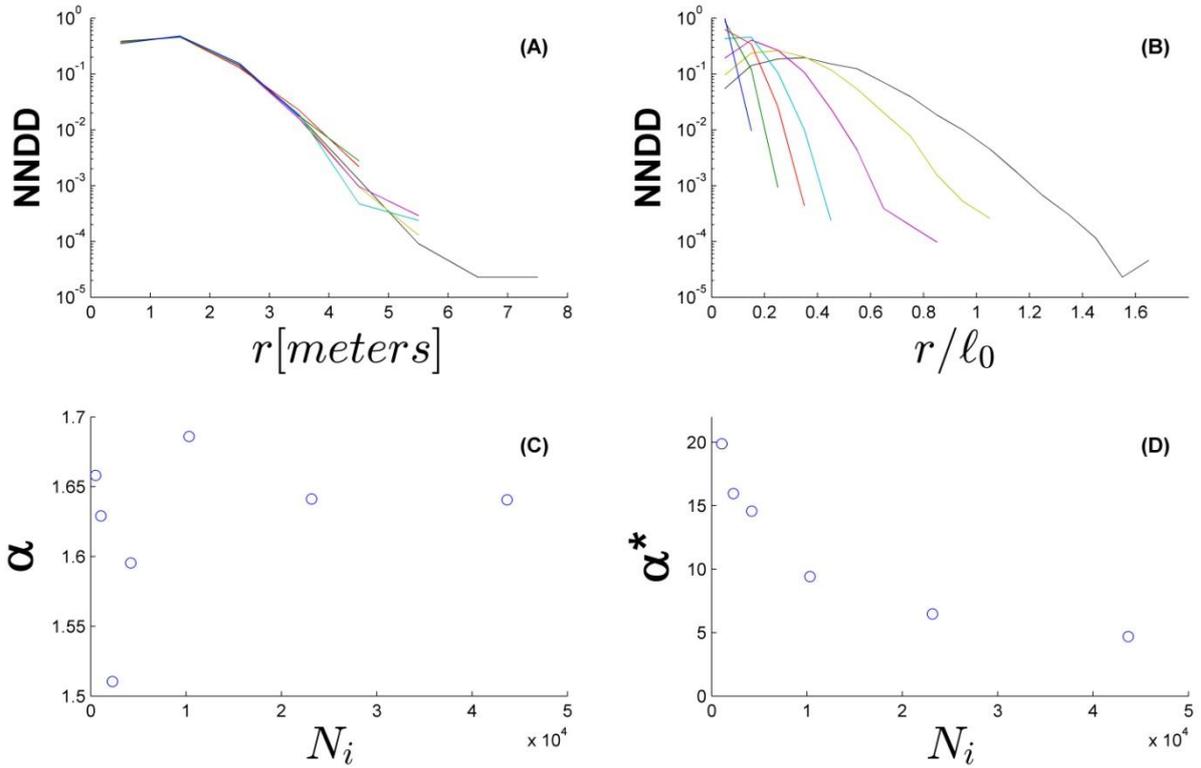

Figure 2.4: Same as fig. 2.1, here for the Random Cantor Set (P=0.75). The algorithm that generates a fractal forest, as described in section 1, yields almost no "singletons". i.e. the nearest neighbor of almost any tree is at a fixed distance. To get results that allow for a reasonable comparison with the real data we have added a weak Poissonian noise to the random Cantor set. This procedure was used only for the NNDD. The corresponding results below, for correlations and cluster statistics, were obtained for a fractal forest without any noise.

## 3. Correlations

The correlation function for a Poisson forest (Figure 3.1) is distance independent, as opposed to the power-law decay observed in empirical data. The MLGK (3.2) shows, like in the NNDD case, a crossover to Poisson at large distances, but have a well-defined order at r=10m and r=0.5$l_0$. The same holds for the Cauchy forest (3.3) and for the fractal forest (3.4).

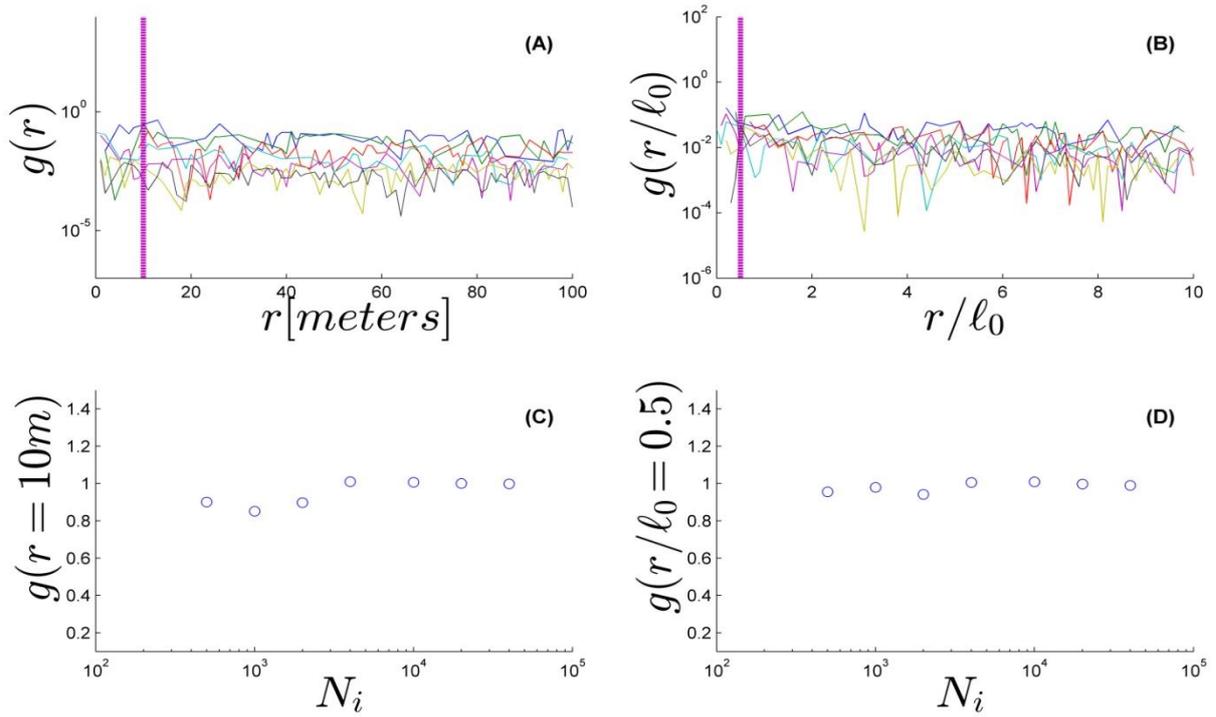

Figure 3.1: The correlation function for a Poisson forest. (A) as a function of real distance $r$. (B) as a function of normalized distance $r/\ell_0$. (C) The height of the correlation function at $r = 10m$ is plotted against the abundance. (D) The height of $g(r/\ell_0 = 0.5)$ vs. $N_i$.

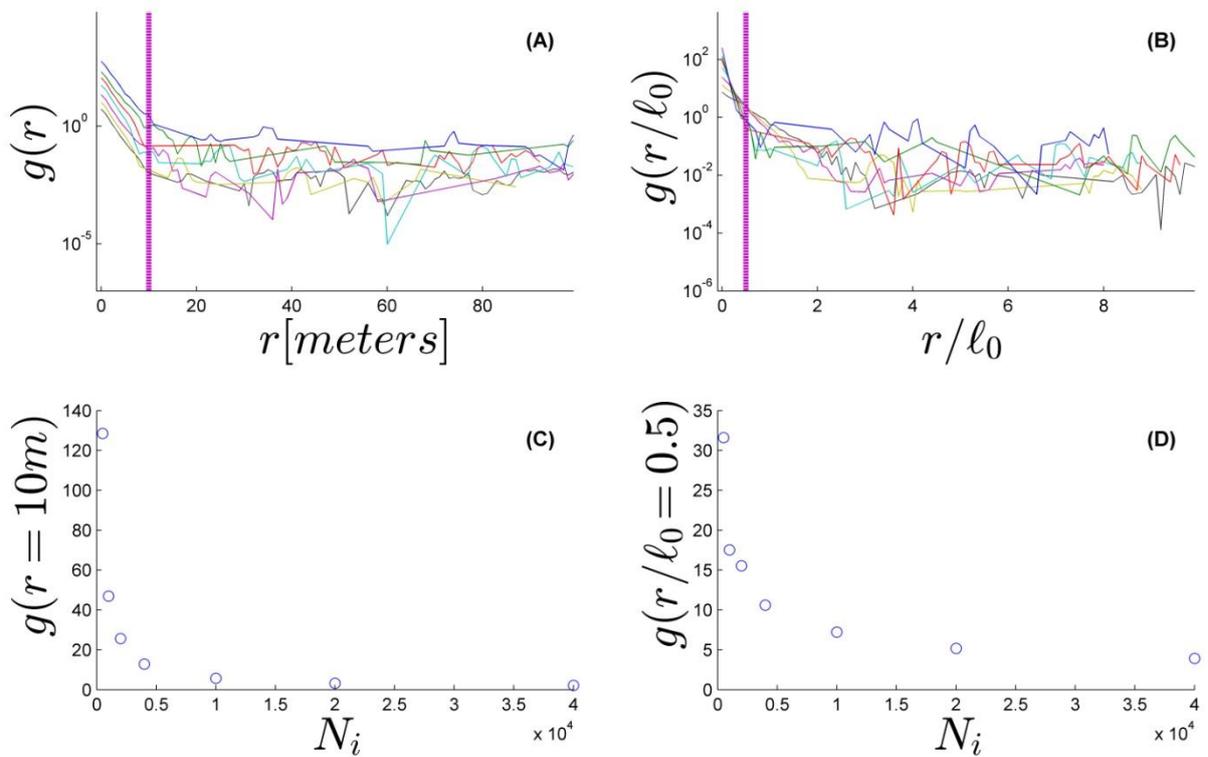

Figure 3.2: Same as fig. 3.1 but for MLGK (μ=0.1).

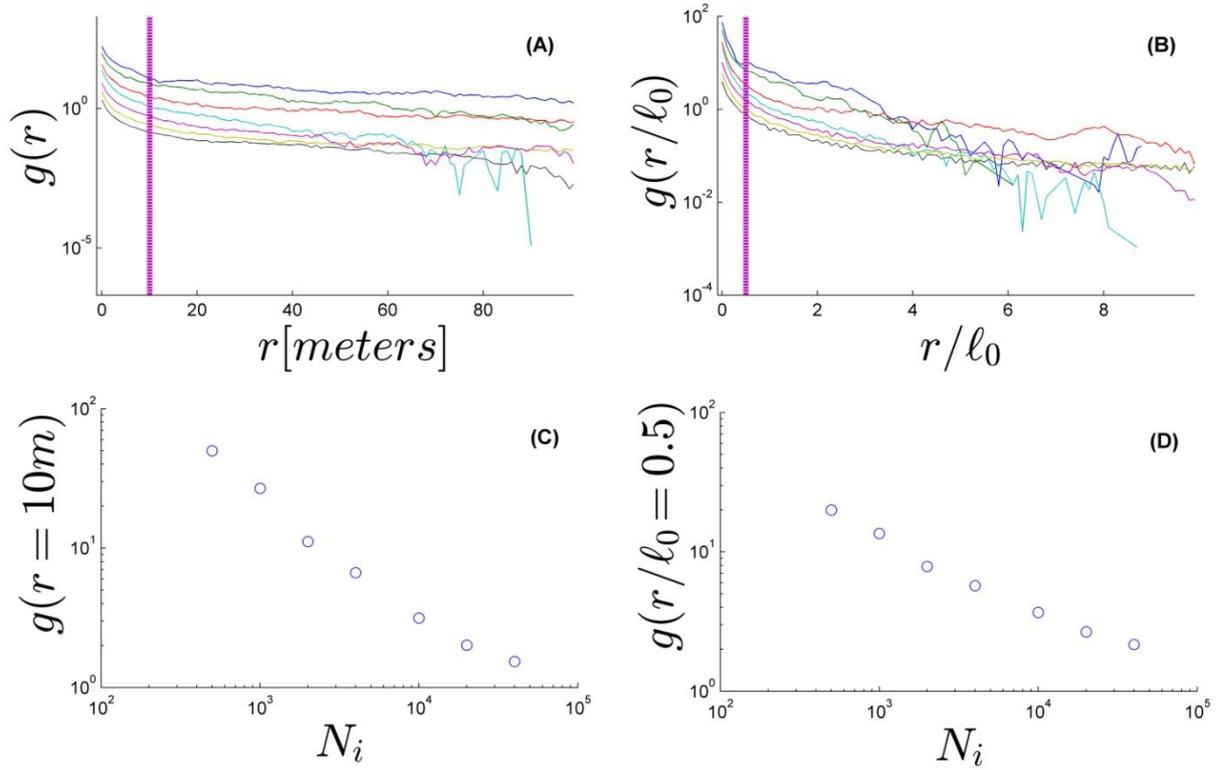

Figure 3.3: Same as fig. 3.1, here for Cauchy kernel (ɣ=20)

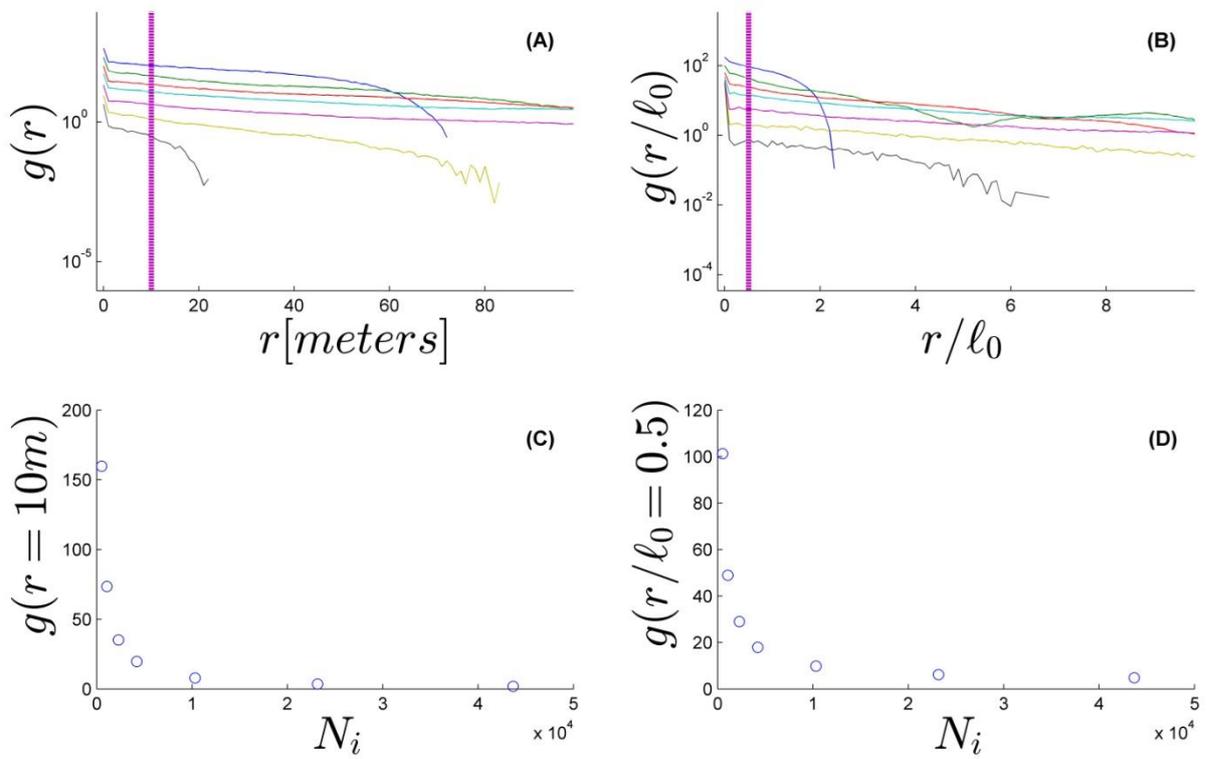

Figure 3.4: Same as fig. 3.1 now for Random Cantor Set (P=0.75).

## 4. Patch statistics at different scales:

A graph showing F($\ell$) vs. $\ell$ has to be abundance independent when the system admit only one length scale. This property holds for the Poisson forest as depicted in Fig. 4.1. Still, the patch statistics in renormalized coordinates for a Poisson forest differs strongly from the empirical results, since the Poisson forest has no real clusters. Figure 4.5 shows together panel (B) of 4.1 and the BCI results from panel (B) of Fig. 3 of the main text, and one can see that there is no overlap between the two clusters/scale graphs.

The fractal forest shows some degree of a collapse in the normalized scales (Fig. 4.2), but the functional dependence on $\ell$ is convex, unlike the concave line that characterizes the real data. In the MLGK (Fig. 4.2) and the Cauchy process (Figure 4.4) there is no collapse at all.

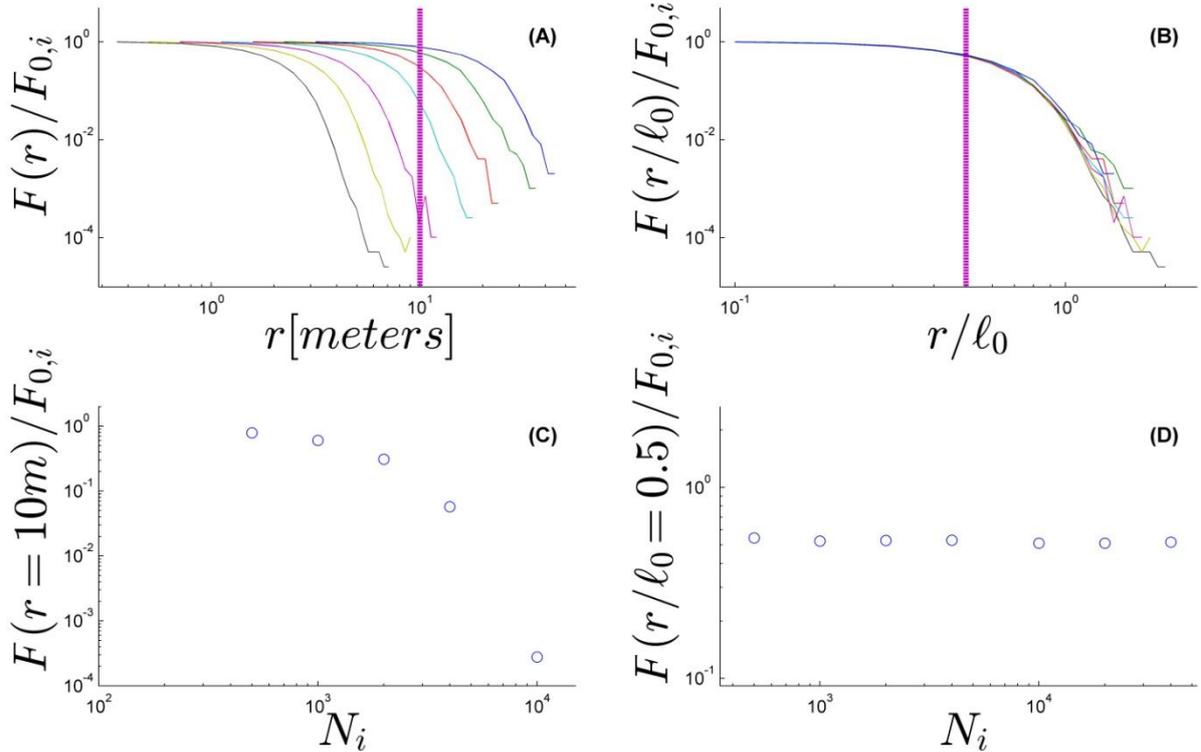

Figure 4.1: Cluster statistics for Poisson distribution. (A) as a function of real distances, r. (B) as a function of normalized distances r/l$_0$. (C) $F(r=10m)/F_{0,i}$ vs. $N_i$ (a cut along the purple dashed line in panel (A)), (D) $F(r/\ell_o=0.5)/F_{0,i}$ vs. $N_i$. As expected, in (D) the correlation with the abundance disappears.

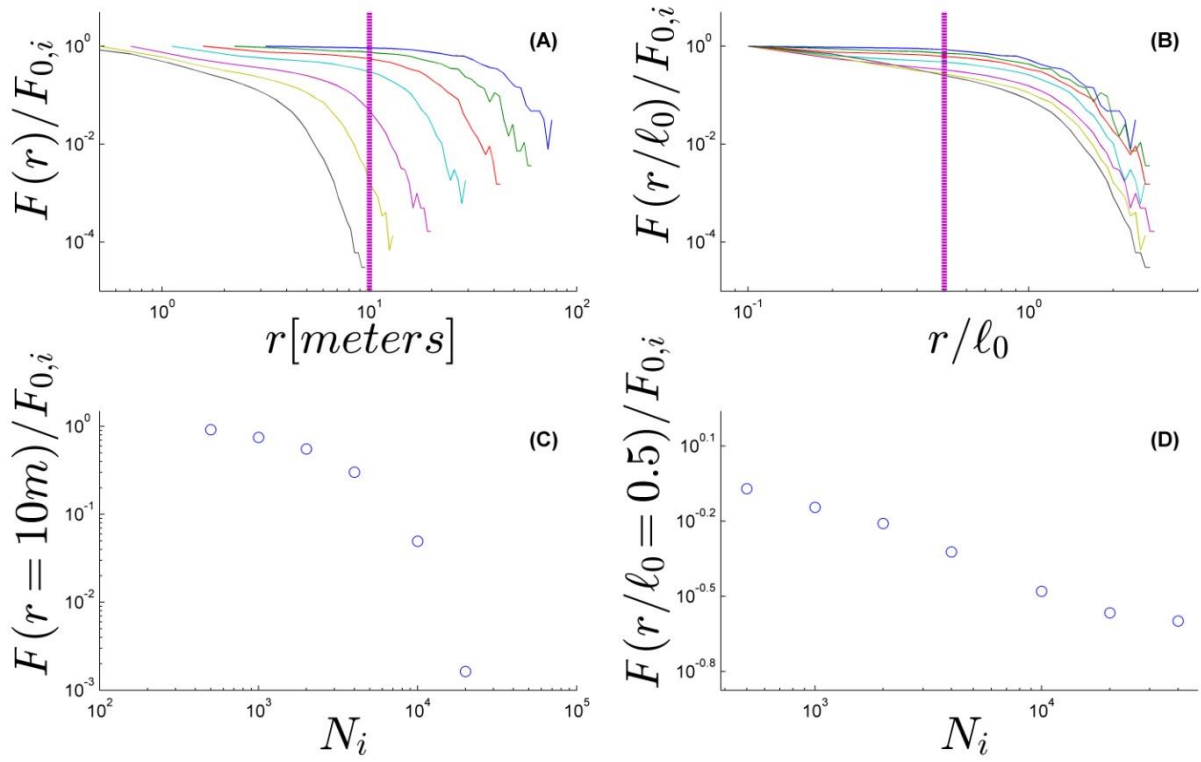

Figure 4.2: Same as fig. 4.1 for MLGK (μ=0.1).

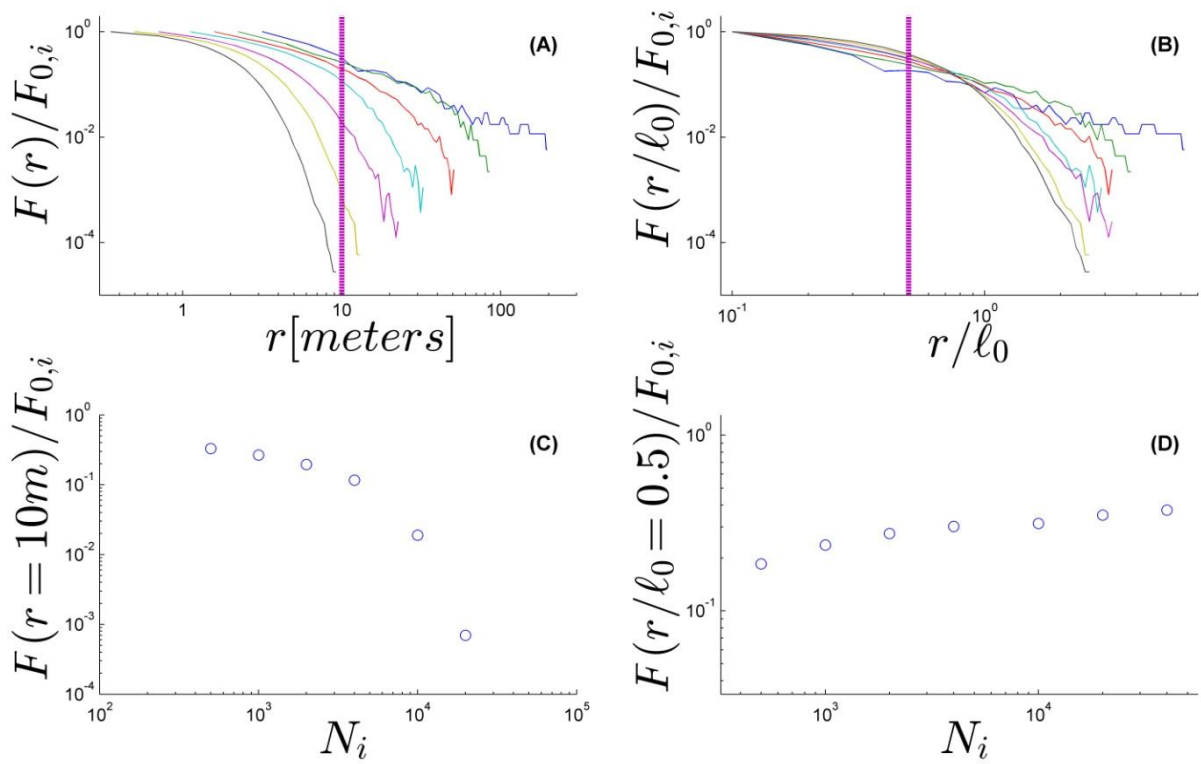

Figure 4.3: Same as fig. 4.1 for Cauchy kernel (ɣ=20)

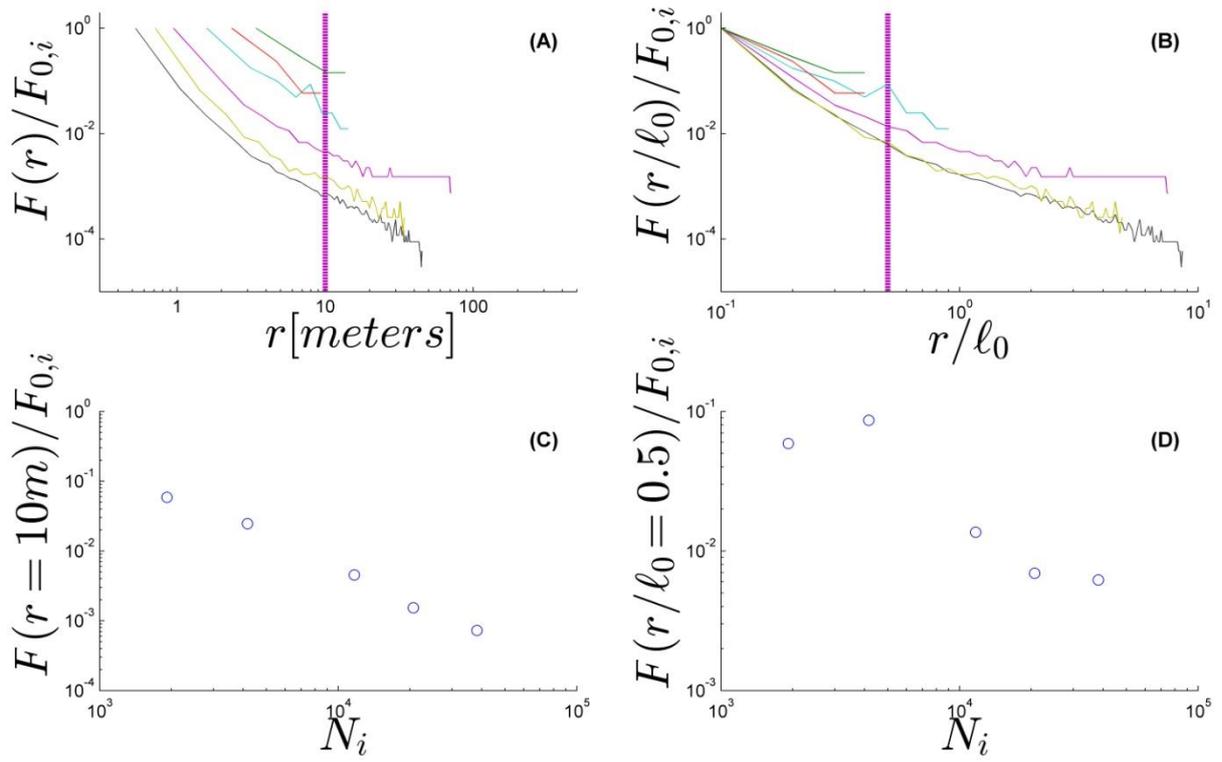

Figure 4.4: Same as fig. 4.1 for Random Cantor Set (P=0.75).

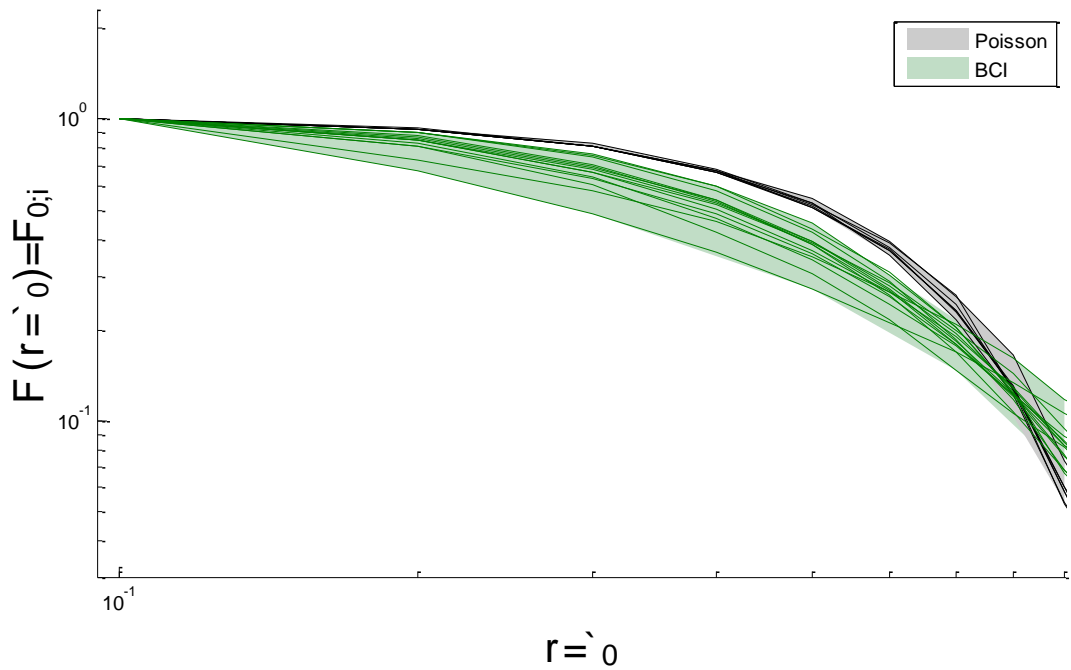

Figure 4.5: $F(r/\ell_o)/F_{0,i}$ for Poisson distribution (black lines) and for BCI species (green lines).

# 5. Exceptional species

These figures show the spatial patterns of three exceptional species in the BCI, see main text. Every red circle corresponds to >1cm individual of the focal species.

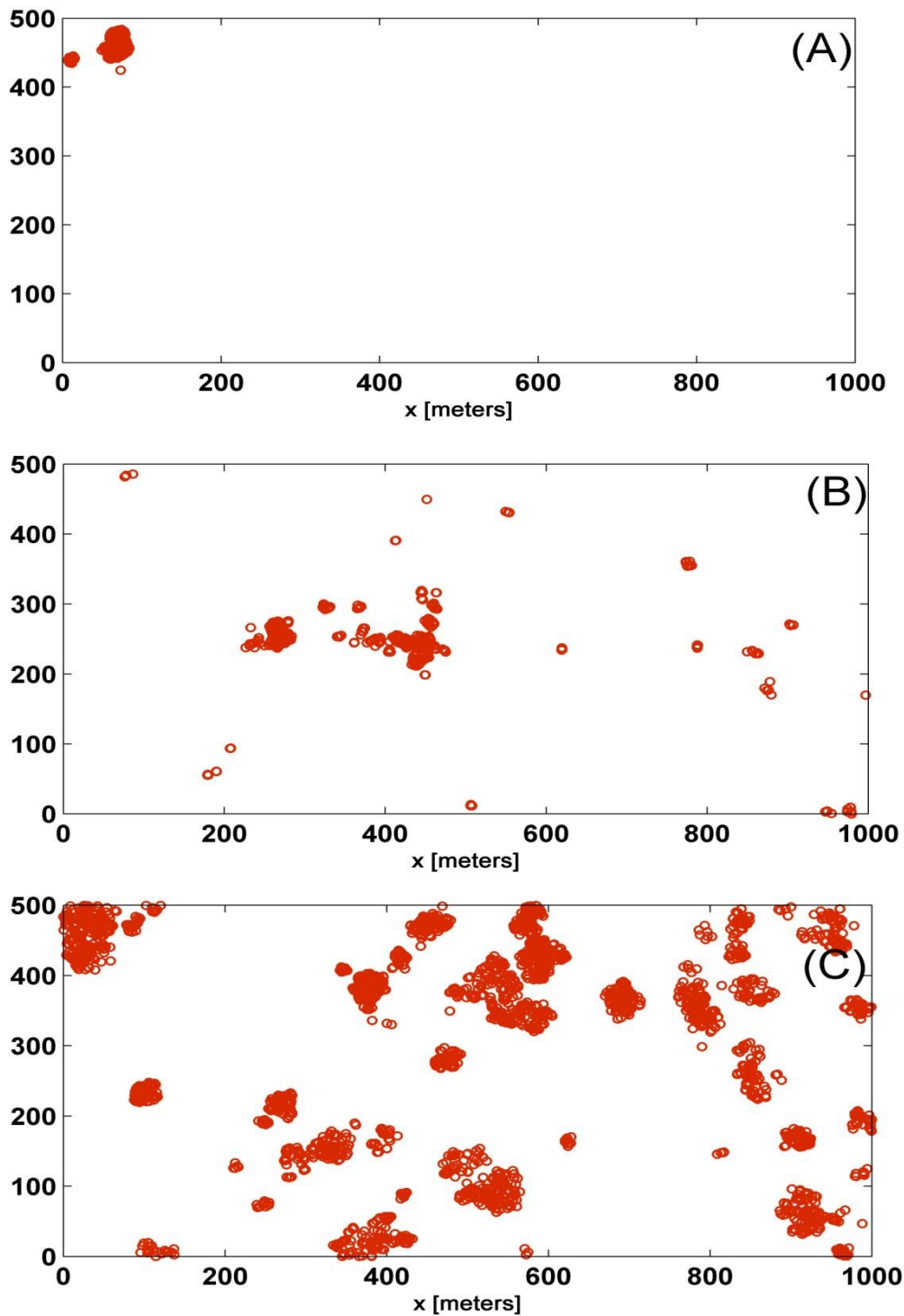

Figure 14: Exceptional species (A) Anaxagorea panamensis (B) Bactris major (C) Rinorea sylvatica.

25